\documentclass[aps, prb, twocolumn, superscriptaddress, 10pt, floatfix]{revtex4-1}
\usepackage{graphicx}
\usepackage[hidelinks]{hyperref}

\begin{document}

\title{Parallel InAs nanowires for Cooper pair splitters with Coulomb repulsion}

\author{Oliv\'er K\"urt\"ossy}
\affiliation{Department of Physics,  Institute of Physics, Budapest University of Technology and Economics, M\H{u}egyetem rkp. 3, H-1111 Budapest, Hungary}

\affiliation{MTA-BME Nanoelectronics Momentum Research Group, M\H{u}egyetem rkp. 3, H-1111 Budapest, Hungary}

\author{Zolt\'an Scher\"ubl}
\affiliation{Department of Physics,  Institute of Physics, Budapest University of Technology and Economics, M\H{u}egyetem rkp. 3, H-1111 Budapest, Hungary}

\affiliation{MTA-BME Nanoelectronics Momentum Research Group, M\H{u}egyetem rkp. 3, H-1111 Budapest, Hungary}

\affiliation{Univ. Grenoble Alpes, CEA, Grenoble INP, IRIG, PHELIQS, 38000 Grenoble, France}

\author{Gerg\H o F\"ul\"op}
\affiliation{Department of Physics,  Institute of Physics, Budapest University of Technology and Economics, M\H{u}egyetem rkp. 3, H-1111 Budapest, Hungary}

\affiliation{MTA-BME Nanoelectronics Momentum Research Group, M\H{u}egyetem rkp. 3, H-1111 Budapest, Hungary}

\author{Istv\'an Endre Luk\'acs}
\affiliation{Center for Energy Research, Institute of Technical Physics and Material Science, Konkoly-Thege Mikl\'os \'ut 29-33., H-1121, Budapest, Hungary}

\author{Thomas Kanne}
\affiliation{Center  for  Quantum  Devices,  Niels  Bohr  Institute, University  of  Copenhagen,  2100  Copenhagen,  Denmark}

\author{Jesper Nyg{\aa}rd}
\affiliation{Center  for  Quantum  Devices,  Niels  Bohr  Institute, University  of  Copenhagen,  2100  Copenhagen,  Denmark}

\author{P\'eter Makk}
\email{makk.peter@ttk.bme.hu}
\affiliation{Department of Physics,  Institute of Physics, Budapest University of Technology and Economics, M\H{u}egyetem rkp. 3, H-1111 Budapest, Hungary}

\affiliation{MTA-BME Correlated van der Waals Structures Momentum Research Group, M\H{u}egyetem rkp. 3, H-1111 Budapest, Hungary}

\author{Szabolcs Csonka}
\email{csonka.szabolcs@ttk.bme.hu}
\affiliation{Department of Physics,  Institute of Physics, Budapest University of Technology and Economics, M\H{u}egyetem rkp. 3, H-1111 Budapest, Hungary}

\affiliation{MTA-BME Nanoelectronics Momentum Research Group, M\H{u}egyetem rkp. 3, H-1111 Budapest, Hungary}

\begin{abstract}
Hybrid nanostructures consisting of two parallel InAs nanowires connected by an epitaxially grown superconductor (SC) shell recently became available. Due to the defect-free SC-semiconductor interface and the two quasi-one-dimensional channels being close by, these novel platforms can be utilized to spatially separate entangled pairs of electrons by using quantum dots (QD) in the so-called Cooper pair splitting (CPS) process. The minimized distance between the QDs overcomes the limitations of single-wire-based geometries and can boost the splitting efficiency. Here we investigate CPS in such a device, for the first time, where strong inter-dot Coulomb repulsion is also present and studied thoroughly. We analyze theoretically the slight reduction of the CPS efficiency imposed by the Coulomb interaction and compare it to the experiments. Despite the competition between crossed Andreev reflection (CAR) and inter-wire capacitance, a significant CPS signal is observed indicating the dominance of the superconducting coupling. Our results demonstrate that the application of parallel InAs nanowires with epitaxial SC is a promising route for the realization of parafermionic states relying on enhanced CAR between the wires.
\end{abstract}

\date{\today}

\pacs{}


\maketitle

\begin{figure}[htp]
\includegraphics[width=0.5\textwidth]{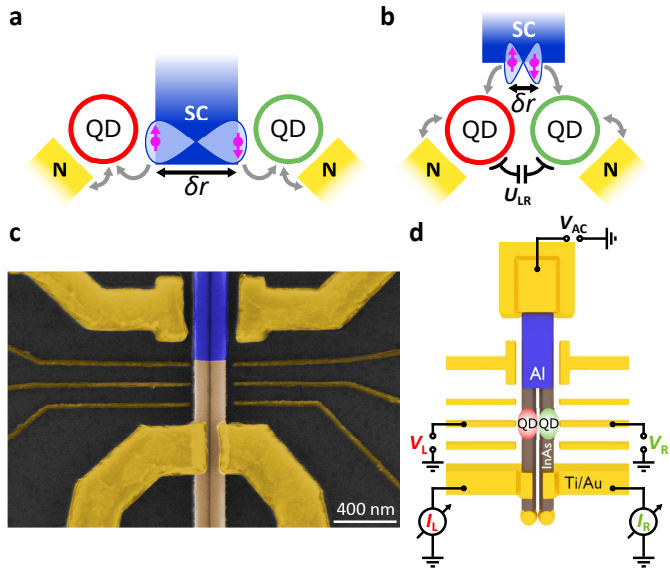}
\caption{\label{device_outline} \textbf{Device outline.} \textbf{a} Concept of a single-nanowire based Cooper pair splitter. Lower bound for $\delta r$ is the width of the SC with negligible cross-capacitance between the QDs. \textbf{b} Same as panel \textbf{a}, but in double, parallel nanowires. While $\delta r$ can be minimized, $U_\mathrm{LR}$ becomes significant. \textbf{c} False-colored scanning electron microscopy (SEM) image of the device. The epitaxial Al (blue) was etched away from half of the InAs nanowires (brown). Two separate and a shared Ti/Au contacts and side gate electrodes (yellow) were evaporated to control the transport. \textbf{d} Schematic illustration of the measurement setup. The left (red) and right (green) QDs were tuned by $V_\mathrm{L}$ and $V_\mathrm{R}$, respectively, while the SC was biased and the currents in the two arms were measured, yielding the differential conductance $G_\mathrm{L}$ and $G_\mathrm{R}$.}
\end{figure}

\textbf{Introduction.} Nowadays superconducting nanostructures built from InAs nanowires attract huge attention in the field of quantum electronics. Besides being a versatile platform to investigate the spatial separation of entangled electrons originating from a SC via QDs, namely CPS,\cite{recher2001andreev,lesovik2001electronic,recher2002superconductor} they are also promising candidates to host such unique systems as Andreev-qubits\cite{zazunov2003andreev,koch2007charge,janvier2015coherent,hays2018direct,tosi2019spin} or topologically protected bound states, e.g. Majorana fermions (MF).\cite{kitaev2001unpaired,lutchyn2010majorana,oreg2010helical,vaitiekenas2020flux,prada2020andreev} Although single-wire circuits have already been realized to study MFs\cite{mourik2012signatures,das2012zero,deng2016majorana,albrecht2016exponential,gul2018ballistic,grivnin2019concomitant} and CPS\cite{hofstetter2009cooper,herrmann2010carbon,hofstetter2011finite,schindele2012near,das2012high,deacon2015cooper}, latest theoretical research even predicts non-Abelian anyons with more exotic statistics, so-called parafermions, to appear in devices built from two parallel nanowires.\cite{klinovaja2014time,reeg2017destructive,thakurathi2018majorana} It is theorized that Z3 parafermions can arise in a pair of tightly placed wires with a joint SC by exploiting the CAR and the Coulomb interaction.\cite{byers1995probing,deutscher2000coupling,lesovik2001electronic} As a natural indication of the strong SC-mediated coupling, one might expect the enhancement of CPS in such an arrangement as the distance ($\delta r$) between the two points where the split electrons are injected into the parallel QDs can be substantially decreased (see Figs. \ref{device_outline}\textbf{a-b}).\cite{recher2001andreev,leijnse2013coupling}

CPS in individual nanowires placed manually next to each other has already been reported.\cite{baba2018cooper,ueda2019dominant} However, novel hybrid nanostructures became available recently, where double InAs nanowires are grown in close vicinity and are connected by an in-situ evaporated SC Al.\cite{kanne2021double,kurtossy2021andreev,vekris2021josephson,vekris2021asymmetric} The latter property can satisfy a set of strict requirements from the geometry: the direct tunneling between the QDs is prevented,\cite{fulop2014local,fulop2015magnetic} and the SC-nanowire interface becomes high-quality,\cite{krogstrup2015epitaxy,chang2015hard} rendering them a suitable platform for CPS circuits. As a result of the minimal distance of QDs preferred for the CPS in such setups, the inter-dot Coulomb repulsion ($U_\mathrm{LR}$) also becomes considerable (see Fig. \ref{device_outline}\textbf{b}). This parasitic effect competes with the SC-induced inter-wire transport processes, where two electrons are transmitted through the adjacent QDs. Although previous theoretical works investigated the impact of the inter-dot capacitance in parallel double QD systems,\cite{eldridge2010superconducting,trocha2015spin,hussein2016double} they focused on the regime of strong coupling to the SC (Andreev limit),\cite{buitelaar2002quantum,sand2007kondo,eichler2007even,grove2009superconductivity,pillet2010andreev,lee2014spin,jellinggaard2016tuning,scherubl2020large,prada2020andreev} the opposite of what is desired for CPS.\cite{recher2001andreev,sauret2004quantum} 

In this paper, we report a significant CPS signal observed in parallel InAs nanowires with an epitaxial Al layer. We discuss the CPS through a parallel double QD system and estimate the relative reduction of its efficiency in the presence of finite inter-dot Coulomb repulsion energies. Despite the strength of this capacitive coupling being comparable to the superconducting gap ($\Delta$) in our experiments, we report a higher splitting efficiency than in most preceding experiments performed in InAs nanowires. Our findings demonstrate that double nanowires with epitaxial SC are ideal for future applications, where the dominance of crossed Andreev processes is needed.\cite{klinovaja2014time,reeg2017destructive,thakurathi2018majorana}

\textbf{Device outline \& setup.} The system studied here is illustrated in Figs. \ref{device_outline}\textbf{b-d}. A pair of parallel InAs nanowires were initially merged by an epitaxial Al covering 2 facets of them (blue), which was partially removed from the nanowires (brown) by using wet etch method (for details, see Appendix).\cite{kanne2021double} Two normal metal electrodes (Ti/Au, yellow) were formed to the etched segments contacting the wires separately, while a third normal electrode was evaporated on the epitaxial SC (see Fig. \ref{device_outline}\textbf{d}) with a distance of $\sim$2$\,\mu$m measured from the border of the etching. We note that this length exceeds the superconducting coherence length of the Al ($\xi$). Additional 3-3 side gate electrodes were defined to create electrostatically a QD in each of the wires, thus obtaining parallel SC-QD-N junctions with a joint SC (N stands for normal metal). Low-temperature electronic transport measurements were carried out at a base temperature of 40 mK. 

Tunnel barriers were formed by adjusting the voltage on the outer side gates. If the coupling of the QDs is stronger to the SC than to the normal leads, sub-gap states can be formed and one enters the Andreev limit.\cite{buitelaar2002quantum,sand2007kondo,eichler2007even,grove2009superconductivity,pillet2010andreev,lee2014spin,jellinggaard2016tuning,scherubl2020large,prada2020andreev} In our case, the tunnel barriers were set such that the opposite limit was reached, where the normal leads were coupled strongly. This allowed the QDs to be emptied rapidly without blocking the transport, hence making it suitable to investigate CPS.\cite{recher2001andreev,sauret2004quantum} The middle gate electrodes served as plunger gates to tune the level position of the left (red) and right (green) QDs by $V_\mathrm{L}$ and $V_\mathrm{R}$, respectively (see Fig. \ref{device_outline}\textbf{d}). Differential conductance $G_\mathrm{L} = \mathrm{d}I_\mathrm{L}/\mathrm{d}V_\mathrm{AC}$ and $G_\mathrm{R} = \mathrm{d}I_\mathrm{R}/\mathrm{d}V_\mathrm{AC}$ through the left and right QDs were recorded on the normal contacts biased by $V_\mathrm{SD}$ applied on home-built I/V converters. With this geometry, we minimized $\delta r$ in order to boost the superconducting inter-dot correlations as suggested in Refs. \onlinecite{recher2001andreev,leijnse2013coupling}, since the CPS current ($\Delta I$) is suppressed as: 
\begin{equation}\label{eq:suppression}
\Delta I\propto \mathrm{exp}\lbrace-2\delta r/\pi\xi\rbrace ,
\end{equation}
however, we also introduced a finite $U_\mathrm{LR}$ at the same time, as shown in Fig. \ref{device_outline}\textbf{b}.

\begin{figure}[htp]
\includegraphics[width=0.5\textwidth]{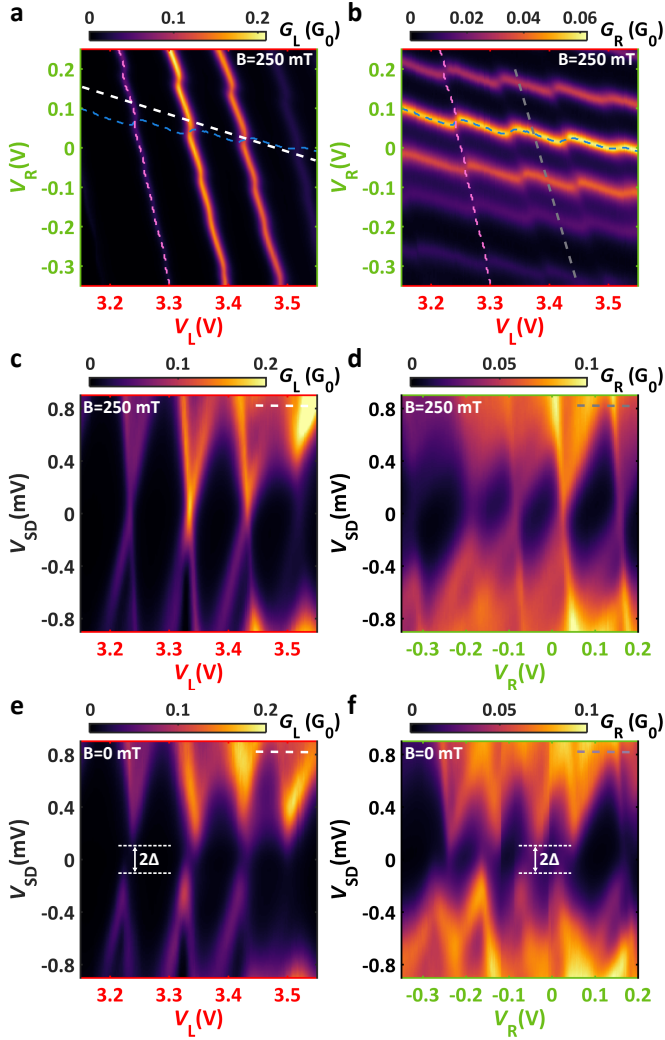}
\caption{\label{diamonds} \textbf{QD characterization.} \textbf{a-b} Zero-bias stability maps of the left and right QDs in the normal state. Finite capacitance between the the plunger gates to the opposite QDs with the strong inter-dot Couloumb repulsion establishes a honeycomb structure in the phase diagram. \textbf{c-d} Bias spectroscopy of the left and right QDs in the normal state and \textbf{e-f} in the superconducting state along the white and gray dashed lines in panels \textbf{a-b}. $U_\mathrm{LR} = 0.15\,$meV was read off from the phase diagram, while charging energies of the left and right QDs were extracted as $U_\mathrm{L} = 0.9\,$meV and $U_\mathrm{R} = 0.7\,$meV, respectively, with $\Delta = 0.15\,$meV from the Coulomb-blockade spectroscopy.}
\end{figure}

\textbf{Results.} In the following, we discuss the spectrum of the created double QD system. Figs. \ref{diamonds}\textbf{a-b} show the zero-bias conductance of the left and right QDs, respectively, as a function of the two plunger gate voltages in the normal state reached by $B=250\,$mT out-of-plane magnetic field. The finite slopes of the lines are attributed to the capacitance between the left (right) plunger gate and the right (left) QD. The resonance lines of each QD shift in the phase diagram when the other one is being charged (or discharged) Due to the significant inter-dot Coulomb repulsion, resulting in an effective gating and exhibiting a honeycomb pattern, which is well-known for capacitively coupled double QDs.\cite{van2002electron} Figs. \ref{diamonds}\textbf{c-d} show normal-state finite-bias spectroscopy of the left and right QDs performed along the white and gray dashed line in Figs. \ref{diamonds}\textbf{a-b}, while Figs. \ref{diamonds}\textbf{e-f} show the same measurements in the superconducting state ($B=0$). The charging energies of the left and right QDs were read off as $U_\mathrm{L} = 0.9\,$meV and $U_\mathrm{R} = 0.7\,$, respectively, while the strength of the inter-dot Coulomb repulsion was found to be $U_\mathrm{LR} = 0.15\,$meV. In Figs. \ref{diamonds}\textbf{e-f} the Coulomb resonances split up at zero bias and a soft gap opens with $2\Delta$ energy in the excitation spectra (with $\Delta = 0.15\,$meV), typical for SC-QD-N junctions. The tip of the diamonds also shift in gate voltage and the lack of sub-gap states confirms the weak coupling towards the SC, needed for CPS experiments. 

\begin{figure}[htp]
\includegraphics[width=0.5\textwidth]{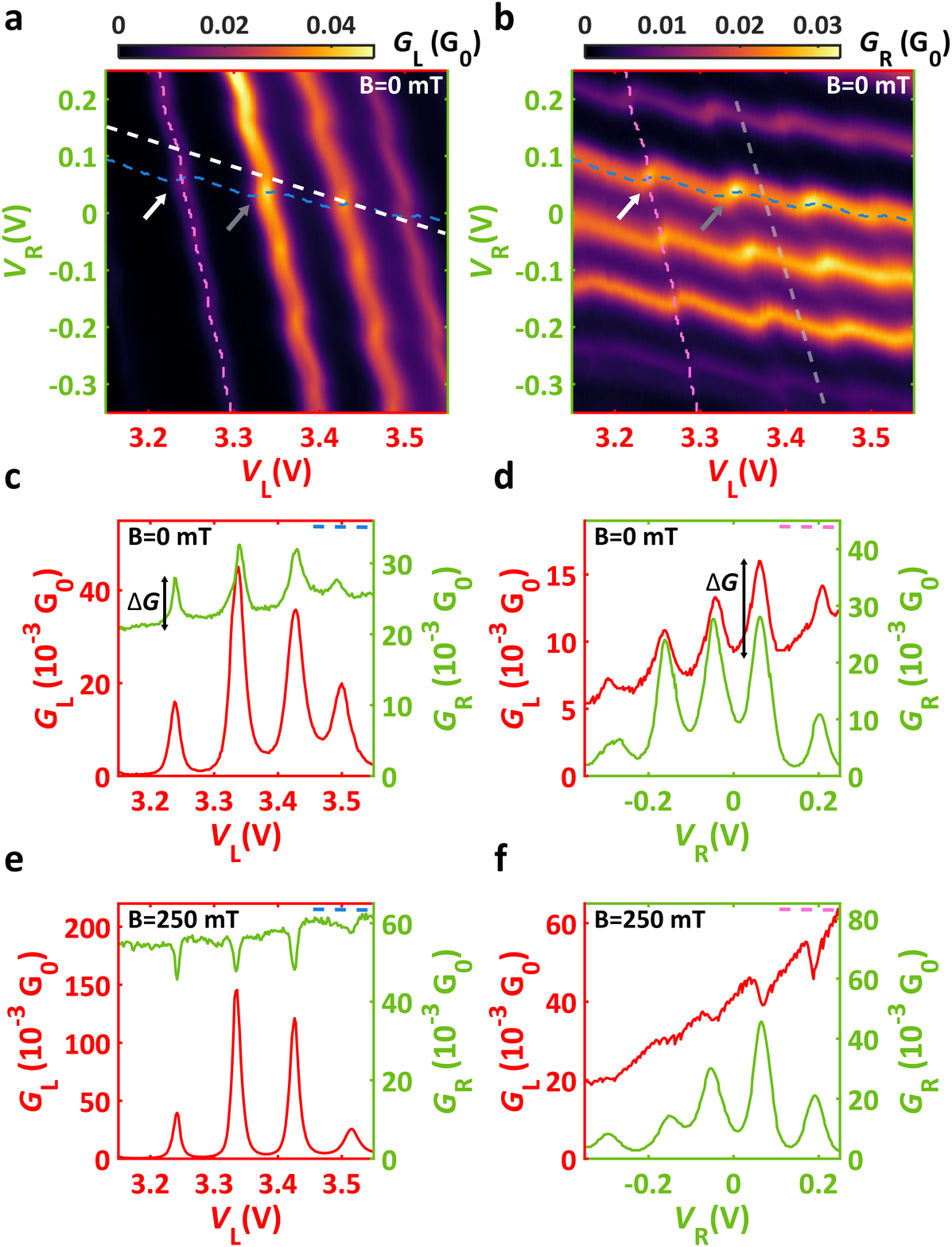}
\caption{\label{cps} \textbf{CPS validation.} \textbf{a-b} Same gate stability maps as in Figs. \ref{diamonds}\textbf{a-b}, but in the superconducting state. \textbf{c} Conductance along a selected resonance of the right QD indicated by the blue dashed line in panels \textbf{a-b}. There is a strong positive correlation between $G_\mathrm{L}$ and $G_\mathrm{R}$. \textbf{d} Similar cut to panel \textbf{c}, but along a resonance of the left QD shown by the pink dashed line in panels \textbf{a-b}. The peaks at $V_\mathrm{L}=3.24\,$V match the one at $V_\mathrm{R}=0.07\,$V in panel \textbf{c} with the same amplitude of $\Delta G$. \textbf{e-f} Normal state data of the corresponding cuts in panels \textbf{c-d} (taken along the pink and blue lines in Figs. \ref{diamonds}\textbf{a-b}). Once the superconductivity is suppressed by external magnetic field, negative correlations overtake the non-local peaks arising from the non-negligible $U_\mathrm{LR}$.}
\end{figure}

Let us now explore the zero-bias conductance of the QDs from Figs. \ref{diamonds}\textbf{a-b}, but in the superconducting state. The corresponding charge stability maps are shown in Figs. \ref{cps}\textbf{a-b}. Here the conductance of both QDs is smaller globally compared to Figs. \ref{diamonds}\textbf{a-b}, but non-zero due to the presence of the soft gap. In contrast to the normal-state data where the capacitive cross-talk yields minima at the intersections of the left and right QD resonances, in the superconducting state, maxima develop instead. These are manifested in Fig. \ref{cps}\textbf{b}. by the yellow spots at the shifting resonance lines. To prove that CPS takes place in our system and to quantify its efficiency, we focus on the evolution of the signal amplitudes along single resonance lines. In Fig. \ref{cps}\textbf{c} $G_\mathrm{L}$ and $G_\mathrm{R}$ are plotted along a resonance of the right QD, whose trace is depicted with the blue dashed line in Figs. \ref{cps}\textbf{a-b}. The red curve shows the resonances of the left QD, and we call this signal "local", whereas the green one shows the changes in the conductance of the right QD as a function of $V_\mathrm{L}$, hence we denote this as the "non-local" signal. Although $G_\mathrm{R}$ is always maximal along this cut, well-pronounced peaks emerge when the left QD is also brought to resonance. Similarly, in Fig. \ref{cps}\textbf{d}, where the roles of QDs are interchanged, $G_\mathrm{L}$ and $G_\mathrm{R}$ are demonstrated along the pink dashed line from Figs. \ref{cps}\textbf{a-b}. Here one can see a significant increase in the left QD signal when the right QD becomes resonant as well. E.g. the non-local peaks at $V_\mathrm{L}=3.24\,$V and $V_\mathrm{R}=0.07\,$V can be recognized in both QD signals in Figs. \ref{cps}\textbf{c-d} with the equal height of $\Delta G$ indicated by the black arrows. Nevertheless, these non-local maxima vanish by switching off the superconductivity as shown in Fig. \ref{cps}\textbf{e-f}, where the same analysis was carried out as in Figs. \ref{cps}\textbf{c-d} but in the normal state along the blue and pink lines in Figs. \ref{diamonds}\textbf{a-b}. In the absence of superconductivity, the overall conductivity increases, and the non-local peaks are replaced by dips. These features are robust along each resonance (for additional data, see Supplementary Notes). We emphasize that the dips are caused by the finite inter-dot capacitance and are much deeper than what is expected from resistive cross-talk introduced in Ref. \onlinecite{hofstetter2009cooper}. Positive correlations of the currents only existing in the superconducting state confirm the presence of CPS, whereas the negative ones in the normal state propose an adverse impact of $U_\mathrm{LR}$ to any CPS related processes. 

The CPS efficiency can be defined as $s = 2\Delta G/(G_\mathrm{L} + G_\mathrm{R})$, while visibility of the non-local signal in the left (right) QD is $\eta_\mathrm{L(R)} = \Delta G/G_\mathrm{L(R)}$, where $\Delta G$ is the non-local signal amplitude equal in the two wires (see Fig. \ref{cps}\textbf{c-d}).\cite{hofstetter2009cooper,schindele2012near,fulop2015magnetic,baba2018cooper} We estimated the maximal and average CPS efficiency as $s_\mathrm{max} =$ 29-36\% (see the white arrow in Figs. \ref{cps}\textbf{a-b}) and $\overline{s}=$ 19-28\% in the investigated gate range, respectively. The maximal visibility was found to be $\eta_\mathrm{L,max} =$ 40-49\% and $\eta_\mathrm{R,max}=$ 29-40\% (see the gray arrow in Figs. \ref{cps}\textbf{a-b}). The lower bounds of the given ranges are derived as $\Delta G$ being measured from the baseline of the resonances. The upper bounds are calculated by considering the relative reduction of the conductance usually reaching 15\% (see Figs. \ref{cps}\textbf{e-f}). In this case, the CPS signals are estimated by the sum of the previously described $\Delta G$ and the expected depth of the dips in the QD with smaller average conductance. With this method, the condition of the non-local signals being equal in $G_\mathrm{L}$ and $G_\mathrm{R}$ is still satisfied. In the following, we explore the relative reduction of these quantities caused by the Coulomb interaction between the QDs.

\textbf{Discussion.} A finite $U_\mathrm{LR}$ is expected to quench the CPS as it penalizes transport processes where both QDs are being charged simultaneously. To confirm this assumption and to quantify this effect, we developed a rate equation model to calculate the relative CPS currents in parallel QDs at different electron occupations. 

\begin{figure*}[htp]
\includegraphics[width=1\textwidth]{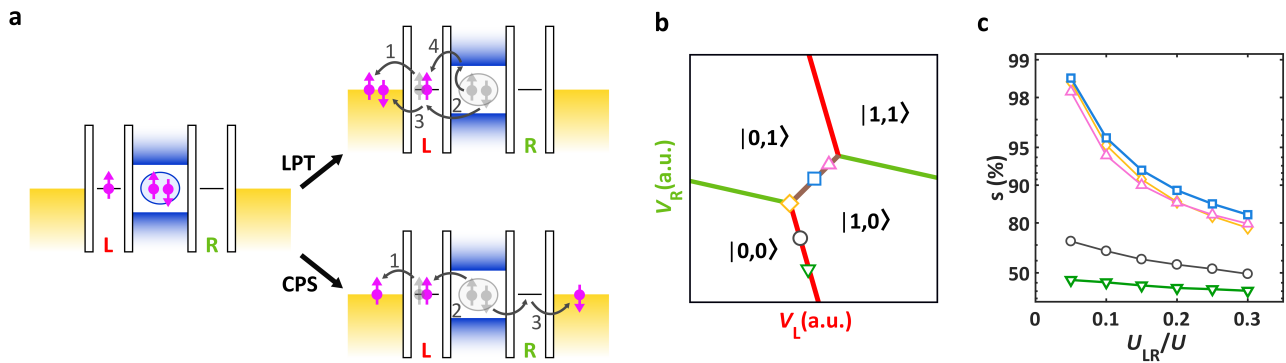}
\caption{\label{theory} \textbf{Transport modeling.} \textbf{a} A possible sequence of LPT and CPS. The arrows with the numbers indicate the event order in the sequence. Electrons in gray symbolize the initial occupations, while the purple ones belong to the final states. \textbf{b} Sketched phase diagram of the honeycomb presenting the border of ground state occupations of both QDs simultaneously. By crossing the red (green) line, the electron number is changed on the left (right) QD. \textbf{c} $U_\mathrm{LR}/U$ dependence of $s$ calculated in different points of the stability diagram, indicated by markers in panel b. While a decreasing tendency of the CPS contribution to the total current can be observed throughout the phase diagram, the efficiency is maximal in the middle of the $|1,0\rangle$ and $|0,1\rangle$ degeneracy lines.}
\end{figure*}

The QDs are treated as single sites in the frame of the Anderson model with $\delta r=0$. We use the notation $|m,n\rangle$ for $|m\rangle_\mathrm{L}\otimes|n\rangle_\mathrm{R}$ describing the ground state electron filling of the left and right QDs, where $m,n=\lbrace 0,\uparrow ,\downarrow,2\rbrace$. The SC and the normal leads are handled by BCS density of states and Fermi distributions, respectively. All tunnel couplings are assumed to be weak compared to the governing energy scales ($U_\mathrm{L(R)}$, $\Delta$) and the transport is entirely described by transition rates determined by 4th order perturbation theory with Fermi's golden rule. The net current in the left (right) lead is obtained by solving the classical master equation in the stationary limit (for further details, see Supplementary Notes).

Among the various transport mechanisms available in the system we distinguish local pair tunneling (LPT) and CPS in the calculations. A sequence of electron tunnelings is treated as LPT if the electrons constituting the Cooper-pair leave the SC to the same QD, while the transport is considered to be CPS if the split electrons exit to separate QDs. One example for each of them is demonstrated in Fig. \ref{theory}\textbf{a} where the order of the events in the sequence is indicated by black arrows and numbers. In both processes the initial and final states have the $|\uparrow ,0\rangle$ character, however, electrons with opposite spins arrive in the same leads at the end of a LPT, while they appear in separate arms in case of CPS (for a detailed discussion see Supplementary Notes).

Intuitively, one expects the CPS efficiency to be maximal when both QDs are close to their resonances.\cite{recher2001andreev} This condition is satisfied in the vicinity of the degeneracy line of the $|1 ,0\rangle$ and $|0 ,1\rangle$ sectors in the phase diagram, i.e. at $-1<\epsilon_\mathrm{L}/U_\mathrm{LR} = \epsilon_\mathrm{R}/U_\mathrm{LR} <0$ on-site energy settings (brown line connecting the triple points in Fig. \ref{theory}\textbf{b}). Therefore we studied the CPS efficiency $s = 2\Delta I/(I_\mathrm{L} + I_\mathrm{R})$ as a function of the inter-dot Coulomb repulsion at different locations in the stability map assigned by the markers in Fig. \ref{theory}\textbf{b}, which are plotted in Fig. \ref{theory}\textbf{c}. For simplicity, the charging energies were chosen as $U_\mathrm{L} = U_\mathrm{R} = U = 1\,$meV and $\Delta = 0.2\,$meV has been used.   

As visible in Fig. \ref{theory}\textbf{c}, $s$ is significantly higher along the degeneracy line of the $|1 ,0\rangle$ and $|0 ,1\rangle$ sectors (pink triangle, blue square, yellow diamond) than anywhere else in the stability diagram and maximal at $\epsilon_\mathrm{L}/U_\mathrm{LR} = \epsilon_\mathrm{R}/U_\mathrm{LR}=-0.5$, i.e. in the middle of the degeneracy line (blue square). This result is consistent with our experimental data where the non-local peaks were positioned to the center of the crossing resonance lines. By moving towards either of the triple points (pink triangle at $\epsilon_\mathrm{L}/U_\mathrm{LR} = \epsilon_\mathrm{R}/U_\mathrm{LR}=-0.8$), $s$ decreases slightly. This small effect originates from the fact any of the CPS cycles that involves both the $|0 ,0\rangle$ and the $|1 ,1\rangle$ configurations as intermediate states (see the CPS process step (1) and (3) depicted in Fig. \ref{theory}\textbf{a}), thereby being penalized by $\sim U_\mathrm{LR}$. Obviously, by increasing $U_\mathrm{LR}$ the triple points separate further, hence suppressing $s$. Once one of the QDs is detuned from the resonance (at the gray circle and the green triangle with fixed $\epsilon_\mathrm{L(R)}$), $s$ drops significantly in accordance with the expectations as LPT starts to dominate the transport. Altogether, $U_\mathrm{LR} \approx\Delta < 0.2\,$meV relevant for our experimental values, the reduction of $s$ does not exceed 10\% compared to the non-interacting case. We note that similar results can be obtained at other on-site energy settings (e.g. in the vicinity of the $|1 ,1\rangle ,|2 ,1\rangle ,|1 ,2\rangle$ and $|2 ,1\rangle ,|1 ,2\rangle ,|2 ,2\rangle$ triple points) due to the symmetry of the charge stability diagram.

As a simple analysis, one can derive the maximal efficiency attainable in single-nanowire-based CPS by assuming the typical values of $\delta r \approx 300\,$nm (see Fig. \ref{device_outline}\textbf{a}) and $U_\mathrm{LR}=0$ (\textbf{I.}), and compare it to the calculation performed $U_\mathrm{LR}=0.15\,$meV by assuming $\delta r=30\,$nm, reasonable for our setup (\textbf{II.}, see Fig. \ref{device_outline}\textbf{a}). According to Eq.$\,$\ref{eq:suppression}, in the former case, $s_{\mathrm{max}}^{\mathrm{I.}}\approx 82$\%, while in the latter one $s_\mathrm{max}^{\mathrm{II.}}\approx 89$\% in principle. By using these values, the geometry exhibiting a minimal $\delta r$, yet a finite $U_\mathrm{LR}$ turns out to be beneficial regarding the CPS efficiency. Naively one can argue it as $U_\mathrm{LR}\propto 1/\delta r$ while $s$ decays exponentially in $\delta r$ as outlined in Eq.$\,$\ref{eq:suppression}. This consideration with the relatively high CPS efficiency reported here in spite of the parasitic inter-dot Coulomb repulsion confirms the advantageous application of parallel InAs nanowires in future SC-semiconductor hybrids.

\textbf{Conclusions.} In summary, we have demonstrated significant Cooper pair splitting signals realized in parallel InAs nanowires connected by an epitaxial Al shell. The behavior of the coupled parallel SC-QD-N junctions was analyzed by spectroscopic measurements in both the superconducting and normal states. Owing to the geometrical properties, strong capacitive interaction was found between the QDs whose effect on the CPS was thoroughly studied. The large CPS efficiency ($s_\mathrm{max}=$ 29-36\%) achieved by the small spatial separation of QDs via the SC proved to outgrow the drawbacks of the inter-dot Coulomb repulsion. The strong crossed Andreev reflection makes the double wires with epitaxial shell promising platforms to develop topological superconducting states, like parafermions.

\section*{Appendix}

\textbf{Device fabrication.} InAs nanowires were grown by MBE in the wurtzite phase along the $\langle 0001\rangle$ direction catalyzed by Au. The pattern of the pre-defined Au droplets allowed to control the geometrical properties of the proposed parallel nanowires, including the diameter, distance, and the corresponding alignment of the cross-sections.\cite{kanne2021double} The 20-nm-thick Al shell (covering 2 facets) was evaporated at low temperature in-situ providing epitaxial, oxide-free layers. The evaporation on such a pair of adjacent nanowires resulted in the merging by the Al. Nanowires with $\sim$80 nm diameter and $\sim$4 $\mu$m length were deposited on a p-doped Si wafer capped with 290 nm thick SiO\textsubscript{2} layer by using an optical transfer microscope with micromanipulators. The Al shell was partially removed by the means of wet chemical etching. A coated MMA/MAA EL-6 double-layer performed as a masking layer, in which designed windows were opened with EBL allowing the MF-321 selective developer to access the Al (45 s). The etching was followed by a careful localization of the wires with high-resolution SEM. The contact electrodes were installed in a separate EBL step with thicker PMMA resist (300 nm), where the sample was exposed to RF Ar milling in the evaporator chamber to remove the native oxide of both the Al and InAs. The process was followed by the metallization of Ti/Au (10/80 nm) with electron beam evaporation without breaking the vacuum. In a second EBL step, the side gate electrodes were created by using thinner PMMA resist (100 nm) and depositing Ti/Au (10/25 nm).

\textbf{Experiments.} Low-temperature characterization was carried out in a Leiden Cryogenics dry dilution refrigerator with a base temperature of 40 mK. Transport measurements were performed with standard lock-in technique by applying 10 $\mu$V AC signal at 137 Hz on the shared SC electrode, whereas the differential conductance of the nanowires was recorded separately via home-built I/V converters. DC bias was adjusted by the offset of the I/V converters. Out-of-plane magnetic field was realized by a superconducting magnet.

\textbf{Author contributions.} O. K. and I. L. fabricated the device, O. K., Z. S. and G. F. performed the measurements and did the data analysis. Z. S. built the theoretical model and developed the numerical simulations. T. K. and J. N. developed the nanowires. All authors discussed the results and worked on the manuscript. P. M. and S. C. proposed the device concept and guided the project.  

\textbf{Acknowledgments.} The authors are thankful to EK MFA for providing their facilities for sample fabrication. We thank D. Olstein, M. Marnauza, A. Vekris, and K. Grove-Rasmussen for experimental assistance, A. P\'alyi, A. Virosztek for discussion, M. G. Beckerne, F. F\"ul\"op, and M. Hajdu for their technical support. This work has received funding Topograph FlagERA, the SuperTop QuantERA network, the FET Open AndQC, the FET Open SuperGate, the COST Nanocohybri network, and from the OTKA K138433 and OTKA FK 132146 grants. P. M. and G. F. acknowledge support from the Bolyai Fellowship. This research was supported by the Ministry of Innovation and Technology and the NKFIH within the Quantum Information National Laboratory of Hungary and by the Quantum Technology National Excellence Program (Project Nr. 2017-1.2.1-NKP-2017-00001), UNKP-21-5 New National Excellence Program of the Ministry for Innovation and Technology from the source of the National Research, Development and Innovation Fund, and the Carlsberg Foundation, and the Danish National Research Foundation. 

\textbf{Competing interests.}
The Authors declare no Competing Financial or Non-Financial Interests.


\end{document}


\renewcommand{\theequation}{S\arabic{equation}}
\renewcommand{\figurename}{Supplementary~Figure}
\renewcommand{\tablename}{Supplementary~Table}
\renewcommand{\thesection}{SUPPLEMENTARY~NOTE~\arabic{section}}
\renewcommand{\thetable}{\arabic{table}}

\title{Supplementary Notes for Parallel InAs nanowires for Cooper pair splitters with Coulomb repulsion}

\author{Oliv\'er K\"urt\"ossy}
\affiliation{Department of Physics,  Institute of Physics, Budapest University of Technology and Economics, M\H{u}egyetem rkp. 3, H-1111 Budapest, Hungary}

\affiliation{MTA-BME Nanoelectronics Momentum Research Group, M\H{u}egyetem rkp. 3, H-1111 Budapest, Hungary}

\author{Zolt\'an Scher\"ubl}
\affiliation{Department of Physics,  Institute of Physics, Budapest University of Technology and Economics, M\H{u}egyetem rkp. 3, H-1111 Budapest, Hungary}

\affiliation{MTA-BME Nanoelectronics Momentum Research Group, M\H{u}egyetem rkp. 3, H-1111 Budapest, Hungary}

\affiliation{Univ. Grenoble Alpes, CEA, Grenoble INP, IRIG, PHELIQS, 38000 Grenoble, France}

\author{Gerg\H o F\"ul\"op}
\affiliation{Department of Physics,  Institute of Physics, Budapest University of Technology and Economics, M\H{u}egyetem rkp. 3, H-1111 Budapest, Hungary}

\affiliation{MTA-BME Nanoelectronics Momentum Research Group, M\H{u}egyetem rkp. 3, H-1111 Budapest, Hungary}

\author{Istv\'an Endre Luk\'acs}
\affiliation{Center for Energy Research, Institute of Technical Physics and Material Science, Konkoly-Thege Mikl\'os \'ut 29-33., H-1121, Budapest, Hungary}

\author{Thomas Kanne}
\affiliation{Center  for  Quantum  Devices,  Niels  Bohr  Institute,University  of  Copenhagen,  2100  Copenhagen,  Denmark}

\author{Jesper Nyg{\aa}rd}
\affiliation{Center  for  Quantum  Devices,  Niels  Bohr  Institute,University  of  Copenhagen,  2100  Copenhagen,  Denmark}

\author{P\'eter Makk}
\email{makk.peter@ttk.bme.hu}
\affiliation{Department of Physics,  Institute of Physics, Budapest University of Technology and Economics, M\H{u}egyetem rkp. 3, H-1111 Budapest, Hungary}

\affiliation{MTA-BME Correlated van der Waals Structures Momentum Research Group, M\H{u}egyetem rkp. 3, H-1111 Budapest, Hungary}

\author{Szabolcs Csonka}
\email{csonka.szabolcs@ttk.bme.hu}
\affiliation{Department of Physics,  Institute of Physics, Budapest University of Technology and Economics, M\H{u}egyetem rkp. 3, H-1111 Budapest, Hungary}

\affiliation{MTA-BME Nanoelectronics Momentum Research Group, M\H{u}egyetem rkp. 3, H-1111 Budapest, Hungary}

\date{\today}
\maketitle

\section{Additional data for CPS}

\begin{figure}[htp]
\includegraphics[width=0.5\textwidth]{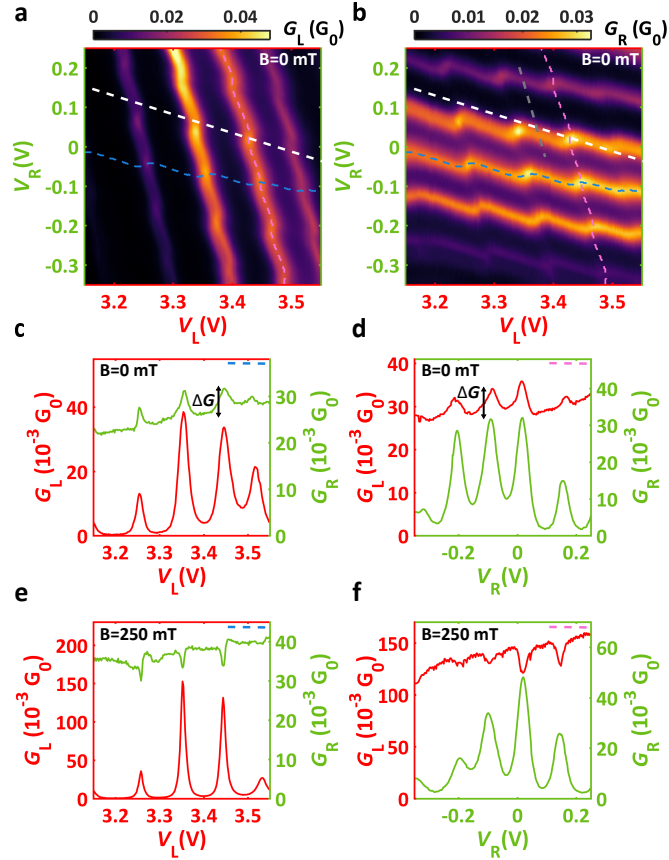}
\caption{\label{additional_data} \textbf{Supplementary analysis of CPS.} \textbf{a-b} Stability map of the left and right QDs in the superconducting state (same as Figs. 3\textbf{a-b}). \textbf{c-d} 1D conductance cuts taken along the blue and pink dashed lines in panel \textbf{a-b}. The non-local signals captured in both left and right QDs at $V_\mathrm{L}=3.45\,$V and $V_\mathrm{R}=-0.08\,$V are measured by $\Delta G$. \textbf{e-f} Same as panels \textbf{c-d}, but in the normal state. The non-local peaks are replaced by significant dips as a result of the finite $U_\mathrm{LR}$.}
\end{figure}

In the main text, the development of CPS signals was analyzed along single resonance lines. Here we provide additional data of other resonances exhibiting qualitatively similar behavior as discussed in Fig. 3, which is shown in Supp Fig. \ref{additional_data}.

Supp Figs. \ref{additional_data}\textbf{c-d} show $G_\mathrm{L}$ and $G_\mathrm{R}$ plotted along the resonance lines of the left and right QDs indicated by the blue and pink dashed lines in Supp Figs. \ref{additional_data}\textbf{a-b}, respectively. While in Supp Fig. \ref{additional_data}\textbf{c} the resonant right QD (green) acquires non-local peaks when the left QD is tuned to resonance too, the same applies in reverse in Supp Fig. \ref{additional_data}\textbf{d}. At $V_\mathrm{L}=3.45\,$V and $V_\mathrm{R}=-0.08\,$V, where the QDs are resonant at the intersection of the blue and pink lines, $\Delta G\approx 8 \,$nA is detected equally in both cuts. In turn, 
these peaks vanish in the normal state and well-pronounced dips evolve instead of them (see Supp Figs. \ref{additional_data}\textbf{e-f} for the same analysis as in panels \textbf{c-d} but without superconductivity). The depth of the dips usually exceeds 10\% of the resonant signal. 

As outlined in the main text, we emphasize that these normal state minima originate from the shifting resonance lines as a result of the capacitive coupling between the QDs. The resistance of each line in the cryostat is $R_\mathrm{W}\approx 220 \, \Omega$ ($1/R_\mathrm{W}\approx 50 \,$G\textsubscript{0}), thus $R_\mathrm{SC}\ll R_\mathrm{W}\ll R_\mathrm{L(R)}$ is satisfied. Consequently, the same circuit model used in Ref. \onlinecite{hofstetter2009cooper} can be applied to describe the classical resistive cross-talk. In the normal state, when the left (right) QD is on resonance, its conductance is expected to decrease ($\Delta G_\mathrm{L(R)}$), when the right (left, $\Delta G_\mathrm{R(L)}$) one also becomes resonant, since a larger current flows through the system thereby reducing the voltage drop on the SC. Quantitatively $\Delta G_\mathrm{L(R)}\approx \Delta G_\mathrm{R(L)}G_\mathrm{L(R)}R_\mathrm{W}$ is estimated, which falls in the order of $<10^{-3}\,$G\textsubscript{0} being negligible compared to the measured depth of the dips.

\section{Capacitive coupling in finite bias measurements}

Previously the Coulomb interaction between the QDs was manifested through the Honey-comb structure in the gate stability maps. Now we briefly present its effect on the evolution of Coulomb diamonds. 

\begin{figure}[htp]
\includegraphics[width=1\textwidth]{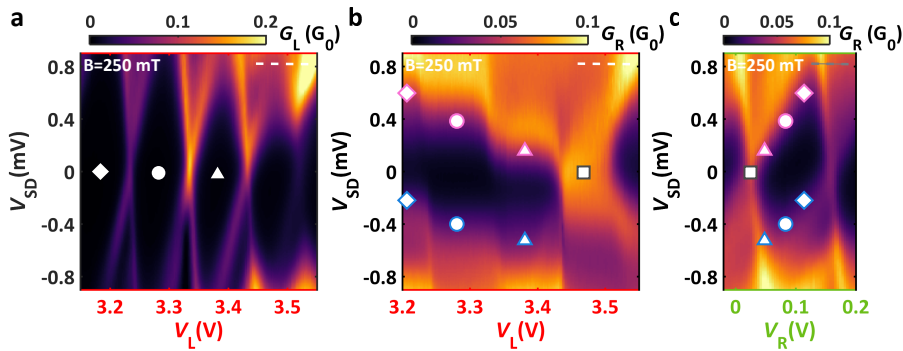}
\caption{\label{coulomb_bias} \textbf{Capacitive coupling in Coulomb-blockade measurements.} \textbf{a-b} Finite-bias spectroscopy along the white dashed line in Supp. Figs. \ref{additional_data}\textbf{a-b} of the left and right QD. In $G_\mathrm{R}$ jumps are induced due to the charging of the left QD. Different charge states are depicted with the triangle, square, and diamond markers. \textbf{c} Coulomb diamond of the state being effectively tuned in panel \textbf{b}. By matching the symbols on the diamond edges from panel \textbf{b}, one can follow the development of the diamond and recognize the asymmetry of the source and drain capacitance relevant for our parallel SC-QD-N junction.}
\end{figure}

Supp Figs. \ref{coulomb_bias}\textbf{a-b} show normal-state finite-bias spectroscopy of the left and right QDs along the white dashed line in Supp. Figs. \ref{additional_data}\textbf{a-b}, parallel to the resonance lines of the right QD. The electron number in the right QD is preserved, however, due to the finite $U_\mathrm{LR}$, $\varepsilon_\mathrm{R}$ is tuned when the left QD is being charged (or discharged). Because of this effective gating, the Coulomb diamond edges of the right QD do not remain on constant energy as one would expect without capacitive coupling, rather they jump along the cut thereby providing a "step-like" evolution of the spectrum. The pink (positive bias) and blue (negative bias) symbols in Supp. Fig. \ref{coulomb_bias}\textbf{b} show these shifts and they match the charge states of the left QD illustrated in Supp. Fig. \ref{coulomb_bias}\textbf{a}. Supp. Fig. \ref{coulomb_bias}\textbf{c} shows bias spectroscopy of the right QD along the gray dashed line in Supp Fig. \ref{additional_data}\textbf{b}. The position of the markers from Supp. Fig. \ref{coulomb_bias}\textbf{b} can be identified and the movement of diamond edges can be tracked exhibiting an asymmetry of the Coulomb diamond as a result of the different source and drain capacitance.

\section{Transport modeling}

\begin{figure}[htp]
\includegraphics[width=0.5\textwidth]{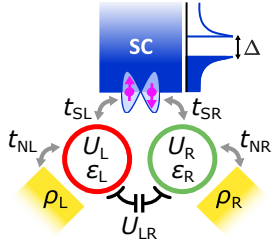}
\caption{\label{model} \textbf{Schematic of the model.} The QDs are modeled as single sites with charging energies $U_\mathrm{L}$, $U_\mathrm{R}$, on-site energies $\varepsilon_\mathrm{L}$, $\varepsilon_\mathrm{R}$, and inter-dot Coulomb repulsion $U_\mathrm{LR}$ between them. Couplings to the SC ($t_\mathrm{SL}$, $t_\mathrm{SR}$) and to the normal leads ($t_\mathrm{NL}$, $t_\mathrm{NR}$) are treated perturbatively. In the calculations, the SC with a gap of $\Delta$ is handled with a BCS density of states, while the leads are considered as Fermi seas with $\rho_\mathrm{L}$, $\rho_\mathrm{R}$.}
\end{figure}

In this section, we discuss the theoretical model that we used to simulate the transport in a Cooper pair splitter. As mentioned in the main text, the QDs were described by the Anderson model as single sites allowing the electron numbers to be 0,1, or 2 in each QD. In our limit where the SC and the normal leads are weakly coupled to the QDs, the transport is entirely expressed by transition rates. The sketch of the model is shown in Supp. Fig. \ref{model}.

The Hamiltonians read as:

\bean H = \underbrace{H_{\text{SC}}+H_{\text{DQD}}}_{H_0}+\underbrace{H_{\text{NL}}+H_{\text{NR}}}_{H_{\text{N}}}+\underbrace{H_{\text{TSL}}+H_{\text{TSR}}+H_{\text{TNL}}+H_{\text{TNR}}}_{H_{\text{T}}}.\eean
Here $H_0$ consists of the BCS mean-field Hamiltonian of the SC and the double-QD Andreson term. $H_\mathrm{N}$ describes the normal leads as Fermi-seas, while $H_\mathrm{TL}$ expresses the tunnel couplings of the QDs to the SC and the normal leads. They are constructed as
\bean H_{\text{SC}} &=& \sum_{k\sigma} \varepsilon_{\text{S}k} c^{\dagger}_{\text{S}k\sigma} c_{\text{S}k\sigma} + \Delta \sum_k \left( c^{\dagger}_{\text{S}k\uparrow} c^{\dagger}_{\text{S}-k\downarrow} + c_{\text{S}-k\downarrow} c_{\text{S}k\uparrow} \right) = \sum_{k\sigma} E_k \gamma^{\dagger}_{k\sigma} \gamma_{k\sigma} \nonumber \\
H_{\text{DQD}} &=& \sum_{\alpha} \left( \sum_\sigma \varepsilon_{\alpha} n_{\alpha\sigma} + U_{\alpha} n_{\alpha\uparrow} n_{\alpha\downarrow} \right) + U_{\text{LR}} n_{\text{L}} n_{\text{R}} \nonumber \\
H_{\text{NL}} &=& \sum_{k\sigma} \varepsilon_{\text{NL}k} c^{\dagger}_{\text{L}k\sigma} c_{\text{L}k\sigma} \nonumber \\
H_{\text{NR}} &=& \sum_{k\sigma} \varepsilon_{\text{NR}k} c^{\dagger}_{\text{R}k\sigma} c_{\text{R}k\sigma} \nonumber \\ 
H_{\text{TSL}} &=& \sum_{k\sigma} t_{\text{SL}} \left( c^{\dagger}_{\text{S}k\sigma} d_{\text{L}\sigma} + d^{\dagger}_{\text{L}\sigma} c_{\text{S}k\sigma} \right) = \sum_{k\sigma} t_{\text{SL}} \left[ \left(u_k \gamma^{\dagger}_{k\sigma} - \sigma v_k \gamma_{-k\bar\sigma}\right) d_{\text{L}\sigma} + d^{\dagger}_{\text{L}\sigma} \left(u_k \gamma_{k\sigma} - \sigma v_k \gamma^{\dagger}_{-k\bar\sigma}\right) \right] \nonumber \\ 
H_{\text{TSR}} &=& \sum_{k\sigma} t_{\text{SR}} \left( c^{\dagger}_{\text{S}k\sigma} d_{\text{R}\sigma} + d^{\dagger}_{\text{R}\sigma} c_{\text{S}k\sigma} \right) = \sum_{k\sigma} t_{\text{SR}} \left[ \left(u_k \gamma^{\dagger}_{k\sigma} - \sigma v_k \gamma_{-k\bar\sigma}\right) d_{\text{R}\sigma} + d^{\dagger}_{\text{R}\sigma} \left(u_k \gamma_{k\sigma} - \sigma v_k \gamma^{\dagger}_{-k\bar\sigma}\right) \right]\nonumber \\ 
H_{\text{TNL}} &=& \sum_{k\sigma} t_{\text{NL}} \left( c^{\dagger}_{\text{L}k\sigma} d_{\text{L}\sigma} + d^{\dagger}_{\text{L}\sigma} c_{\text{L}k\sigma} \right) \nonumber \\ 
H_{\text{TNR}} &=& \sum_{k\sigma} t_{\text{NR}} \left( c^{\dagger}_{\text{R}k\sigma} d_{\text{R}\sigma} + d^{\dagger}_{\text{R}\sigma} c_{\text{R}k\sigma} \right) . \eean
Here $n_{\alpha\sigma} = d^{\dagger}_{\alpha\sigma} d_{\alpha\sigma}$ is the number of electrons with spin $\sigma$ on QD$_{\alpha}$, with $d^{(\dagger)}_{\alpha\sigma}$ being the annihilation (creation) operator of electrons with spin $\sigma$ on QD$_{\alpha}$ and $\alpha=\text{L,R}$ denotes the left, right QD. Similarly, $c^{(\dagger)}_{\mathrm{S}k\sigma}$ and $c^{(\dagger)}_{\mathrm{L(R)}k\sigma}$ are the annihilation (creation) operator of electrons on the SC and on the left (right) Fermi-seas with momentum $k$ and spin $\sigma$, respectively. $t_\mathrm{SL(R)}$ and $t_\mathrm{NL(R)}$ are the hopping amplitude of the left (right) QD to the SC and the left (right) QD to the left (right) normal leads, respectively. In the calculations $U_\mathrm{L}=U_\mathrm{R} = 1\,$meV, and $\Delta = 0.15\,$meV were used roughly matching the experimental values. All hopping amplitudes were chosen as $t_\mathrm{SL(R)}=t_\mathrm{NL(R)}=0.01 \,$meV. By applying a Bogoliubov-transformation:

\bean
c_{k\sigma}= u_k\gamma_{k\sigma} - \sigma v_k\gamma^{\dagger}_{k\bar\sigma} ,
\eean
where $|u_k|^2+|v_k|^2=1$, $H_0$ becomes diagonal.

To obtain the current driven from the SC towards the leads, we are interested in sequences of events, where 2 extra electrons are created in any of the normal leads with momentum $k$ and $k'$ at the end of the cycle. Formally, transitions between the eigenstates of $H_0+H_\mathrm{N}$ are induced by $H_\mathrm{T}$. However, various transport processes are allowed depending on the initial QD occupations, i.e. the on-site energy settings, $\varepsilon_\mathrm{R}, \varepsilon_\mathrm{L}$, therefore one can extend the notation  $|m,n\rangle = |m\rangle_\mathrm{L}\otimes|n\rangle_\mathrm{R}$ introduced in the main text to $|j,k,l,m,n\rangle = |j\rangle_\mathrm{NL}\otimes|k\rangle_\mathrm{L}\otimes|l\rangle_\mathrm{SC}\otimes|m\rangle_\mathrm{R}\otimes|n\rangle_\mathrm{NR}$ where $j,n=\lbrace FS,\uparrow ,\downarrow , 2\rbrace$, $k,m=\lbrace 0,\uparrow ,\downarrow , 2\rbrace$, and $l=\lbrace GS,\uparrow ,\downarrow ,2\rbrace$ ($FS$ and $GS$ refer to the Fermi-sea and the BCS ground states with no quasi-particles, respectively). With this labeling one can track electron occupations and electron movements involving all the SC ($|l\rangle_\mathrm{SC}$), QDs ($|k\rangle_\mathrm{L}$, $|m\rangle_\mathrm{R}$) and normal leads ($|j\rangle_\mathrm{NL}$, $|n\rangle_\mathrm{NR}$).

As an example, let us examine the relevant transport processes starting from the $\ket{FS,\uparrow,GS,0,FS}$ state along the degeneracy line of $|1,0\rangle$ and $|0,1\rangle$ sectors (marking the QD occupations, analyzed in the main text). In this case, 8 final states are accessible with different spin configurations distributed on the QDs and on the leads, which are listed in Supp. Table \ref{CPS_list}. These states can be categorized either as a result of LPT or CPS relying on whether the split electrons leave the SC towards the same or separate QDs, respectively. 3 precedent processes titled as CPS with distinct outcomes are demonstrated in Supp Fig. \ref{CPS_illustration}.

\begin{minipage}{0.33\textwidth}
\begin{table}[H]
    \centering
    \begin{tabular}{| c | c |} \hline
final state & process \\ \hline
$\ket{2,\uparrow,GS,0,FS}$ & LPT \\
$\ket{FS,\uparrow,GS,0,2}$ & LPT \\
$\ket{\uparrow,\uparrow,GS,0,\downarrow}$ & CPS (\textbf{I.}) \\
$\ket{\downarrow,\uparrow,GS,0,\uparrow}$ & CPS \\
$\ket{\uparrow,0,GS,\uparrow,\downarrow}$ & CPS \\
$\ket{2,0,GS,\uparrow,FS}$ & CPS ("singlet", \textbf{II.}) \\
$\ket{\uparrow , \downarrow ,GS,0,\uparrow}$ & CPS ("triplet", \textbf{III.}) \\
$\ket{\uparrow,0,GS,\downarrow,\uparrow}$ & CPS ("triplet") \\
\hline
\end{tabular}
    \caption{\label{CPS_list} \textbf{Table of possible transport processes.} List of possible final states of a transport cycle starting from $\ket{FS,\uparrow,GS,0,FS}$. 3 sequences titled as CPS with different final spin configurations are depicted in Supp. Fig. \ref{CPS_illustration}.}
\end{table}
\end{minipage}
\hfill
\begin{minipage}{0.6\textwidth}
\begin{figure}[H]
    \centering
    \includegraphics[width=1\textwidth]{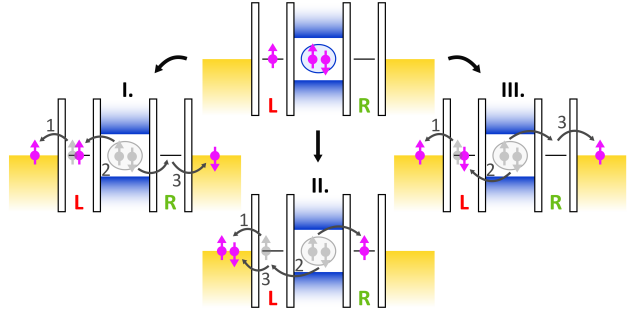} 
    \caption{\label{CPS_illustration} \textbf{CPS processes.} Transport cycles with the same initial, but different final states. Though all 3 are considered as CPS, in scenario \textbf{II.} in a common lead a spin singlet, in scenario \textbf{III.} in separate leads a triplet is created.} 
\end{figure}
\end{minipage}
\bigbreak

In all depicted transport cycles, the sequence starts with a single electron jumping from the left QD to the normal lead followed by splitting a Cooper-pair, but the final states deviate due to the different spin tunnelings. In scenario \textbf{I.}, both the initial and final states have the $|\uparrow ,0$ characters (marking the QD occupations), and electrons with opposite spins appear in separate leads at the end of the cycle, as in a regular CPS (this process was also introduced in Fig. 4. in the main text). In scenario \textbf{II.}, the split electrons tunneling to the QDs have the opposite spins compared to the one in \textbf{I.}. In this particular case, the opposite spins in the left arm originate from different Cooper pairs unlike in LPT and the final state is $\ket{\uparrow\downarrow,0,GS,\uparrow,FS}$. We note that despite a singlet-like state being formed in the left lead, the right QD with an electron stuck there can be emptied in the following transport cycle effectively providing an ordinary CPS. Scenario \textbf{III.} resembles to \textbf{II.}, nevertheless, double tunneling occurs on the right QD instead of the left one bringing the system into $\ket{\uparrow,\downarrow,GS,0,\uparrow}$. Surprisingly, two electrons with the same spins are established in the two Fermi seas, but the transport is still considered as CPS regarding the electrons tunnel to separate QDs from the SC since in the experiments we only measure the currents.

Once the corresponding transport cycles are sorted out, one can calculate the transition rates between the initial and final states in question. Regarding the calculation, technically two different schemes are possible: either both QDs are in blockade, or at least one of them is in resonance with the normal lead. In the former case, all intermediate states are virtual and the transport mechanisms are 4th order processes. Transition rates can be expressed by 4th order perturbation theory using Fermi's golden rule: 

\bean \label{eq:w4} w_{fi}^{(4)} &=& \frac{2\pi}{\hbar} \left| \sum_{pqr}\frac{\bra{f}H_{\text{T}}\ket{r}\bra{r}H_{\text{T}}\ket{q}\bra{q}H_{\text{T}}\ket{p}\bra{p}H_{\text{T}}\ket{i}}{\left( E_i-E_p\right)\left( E_i-E_q\right)\left( E_i-E_r\right)} \right|^2 \delta\left( E_f-E_i\right) ,\eean 
where $i$ and $f$ stand for the initial and final states, respectively, and $E_\alpha$ denotes the total energy of the state $|\alpha\rangle$ ($\alpha =\lbrace i,j,p,q,r\rbrace$). 

However, in the latter case, e.g. in the triple points, the transport cycles are composed by a sequential combination of 1st and 3rd order elementary processes:

\bean \label{eq:w1-3} w_{fi}^{(3)} &=& \frac{2\pi}{\hbar} \left| \sum_{pq}\frac{\bra{f}H_{\text{T}}\ket{q}\bra{q}H_{\text{T}}\ket{p}\bra{p}H_{\text{T}}\ket{i}}{\left( E_i-E_p\right)\left( E_i-E_q\right)} \right|^2 \delta\left( E_f-E_i\right) , \ \ w_{fi}^{(1)}= \frac{2\pi}{\hbar} \left|\bra{f}H_{\text{T}}\ket{i}\right|^2 \delta\left( E_f-E_i\right) . \eean 
Processes depicted in Supp. Fig. \ref{CPS_illustration} are such examples, where the first electron tunneling from the left QD to the lead (1) is a 1st order process. The pair splitting inside the SC and the rest of the tunneling events (2, 3) make up a 3rd order process, and the transition rate in question becomes the sum of $w_{fi}^{(1)}$ and $w_{fi}^{(3)}$. So far we only considered the transition rates of creating electrons with a specific $k$ and $k'$ momentum. To obtain the total transition rates one has to sum up for all possible states available in the normal leads. In a 1st or 3rd order process one can derive them as:

\bean W_{fi}^{(1-3)} =  \sum_{k}w_{fi}^{(1-3)}= \int d\xi \rho_\mathrm{L(R)} (\xi ) w_{fi}^{(1-3)}(\xi ), \eean
where $\rho_\mathrm{L(R)}$ is the density of states of the left (right) lead considered to be constant ($T=0$ temperature is assumed). According to Eq. \ref{eq:w1-3}, $w_{fi}^{(1-3)}\propto\delta (E_f-E_i)$, therefore

\bean W_{fi}^{(1-3)} \propto \int d\xi \rho_\mathrm{L(R)} (\xi ) \delta (\xi ) = \rho_\mathrm{L(R)}(0). \eean
Similarly, in a 4th order process (see Eq. \ref{eq:w4}) the total transition rates read as:

\bean W_{fi}^{(4)} =  \sum_{k k'}w_{fi}^{(4)}= \int d\xi \rho_\alpha  (\xi )\int d\xi ' \rho_\beta  (\xi ')w_{fi}^{(4)}(\xi ,\xi '), \eean
where $\alpha ,\beta =\lbrace \mathrm{L}, \mathrm{R} \rbrace$. 4th order transition rates of any LPT or CPS creating electrons in the leads with momentum $k$ and $k'$ are proportional to a small bias, which was assumed to be $V_\mathrm{AC}=10 \, \mu$V, in accordance with the experiments:

\bean W_{fi}^{(4)} \propto  \int d\xi \rho_\alpha  (\xi )  \int d\xi ' \rho_\beta  (\xi ')  \delta (\xi +\xi ' ) = \int d\xi \rho_\alpha  (\xi ) \rho_\beta  (-\xi )= \rho_\alpha (0)\rho_\beta (0)V_\mathrm{AC}. \eean

Based on the corresponding transition rates, one can solve the classical master equation in the stationary limit:

\bean \dot{P_{q}}=0=\sum_{p \neq q}W_{qp}P_{p}-W_{pq}P_{q}, \eean
where $P_q$ is the occupation of the state $|q\rangle$ with the constriction of $\sum_{q} P_q =1$. As LPT creates 2 electrons in either the left or the right lead and CPS does 1-1 in both of them, the total current in the left (right) normal lead is:
\bnen I_\mathrm{L(R)} = \frac{e}{\hbar} \sum_{pq} \left( 2\cdot W_{qp}^\mathrm{LPT,L(R)} + 0\cdot W_{qp}^\mathrm{LPT,R(L)} + 1\cdot W_{qp}^\mathrm{CPS}\right) P_{p}. \eden
Here $W_{qp}^\mathrm{LPT,L(R)}$ and $W_{qp}^\mathrm{CPS}$ include the transition rates of all LPT and CPS processes creating electrons in the left (right) leads, respectively, and the summation goes over all possible initial and final states.
